# Powerlaw optical conductivity with a constant phase angle in high $T_c$ superconductors.

D. van der Marel*†, H. J. A. Molegraaf*†, J. Zaanen‡, Z. Nussinov‡§, F. Carbone*†, A. Damascelli||¶, H. Eisaki||#, M. Greven||, P. H. Kes‡, & M. Li‡

*Materials Science Centre, University of Groningen, 9747 AG Groningen, The Netherlands

‡Leiden Institute of Physics, Leiden University, 2300 RA Leiden, The Netherlands

|| Department of Applied Physics and Stanford Synchrotron Radiation Laboratory, Stanford University, California 94305, USA

†Present address: Département de Physique de la Matière Condensée, Université de Genève, CH-1211 Genève 4, Switzerland

§Present address: Los Alamos National Laboratories, Los Alamos, NM 87545, USA

¶Present address: Department of Physics and Astronomy, University of British Columbia, Vancouver, British Columbia, V6T 1Z1 Canada

#Present address: Low-Temperature Physics Group, National Institute of Advanced Industrial Science and Technology, Umezono, Tsukuba, 305-8568, Japan

In certain materials with strong electron correlations a quantum phase transition (QPT) at zero temperature can occur, in the proximity of which a quantum critical state of matter has been anticipated[1,2]. This possibility has recently attracted much attention because the response of such a state of matter is expected to follow universal patterns defined by the quantum mechanical nature of the fluctuations. Foremention universality manifests itself through power-law behaviours of the response functions. Candidates are found both in heavy fermion systems[3] and in the cuprate high $T_c$ superconductors[4]. Although there are indications for quantum criticality in the cuprate superconductors[4], the reality and the physical nature of such a QPT are still under debate[5-7]. Here we identify a universal behaviour of the phase angle of the frequency dependent conductivity that is characteristic of the quantum critical region. We demonstrate that the experimentally measured phase angle agrees precisely with the exponent of the optical conductivity. This points towards a QPT in the cuprates close to optimal doping, although of an unconventional kind.

Quantum criticality is associated with a system composed of a (nearly) infinite number of interacting quantum degrees of freedom at zero temperature and it implies, that the system looks on average the same, regardless of the time- and length scale on which it is observed. Electrons on the atomic scale do not exhibit such symmetry, which can only be generated as a collective phenomenon through the interactions between a large number of such entities. In the quantum theory of collective fields one anticipates order at small coupling constant, and for increasing coupling one expects at some point a phase transition to a quantum-disordered state. Quantum criticality in the cuprates, if it exists, occurs as a function of charge carrier doping $x$, at a particular doping level $x_c$ close to where the superconducting phase transition temperature reaches its maximum value. When this phase transition is continuous, a critical state is realized right at the transition, which is characterized by aforementioned scale invariance up to some (non-universal) high-energy cutoff $\Omega$.

The optical conductivity, $\sigma(\omega)=\sigma_1(\omega)+i\sigma_2(\omega)$, is the absorptive ($\sigma_1$) and reactive ($\sigma_2$) current response to a time-varying external electrical field of frequency $\omega$, and is usually expressed as the current-current correlation function, $\chi_{ij}(\tau_1,\tau_2)=<\mathbf{j}(\tau_1),\mathbf{j}(\tau_2)>$, by the Kubo formula. In Fig. 1 we present the experimental optical conductivity function $\sigma_1(\omega)$ of an optimally doped $Bi_2Sr_2Ca_{0.92}Y_{0.08}Cu_2O_{8+\delta}$ single crystal ($T_c$=96 K, work in preparation, H.E., N.Kaneko, D.L.Feng, A.D, P.K.Mang, K.M.Shen, Z-X.Shen&M.G.). In order to facilitate comparison with earlier publications[10-12] we also present $1/\tau(\omega)$ for a number of temperatures, adopting $\omega_p/2\pi c$ = 19364 cm$^{-1}$ for the plasma-frequency. The scattering rate $1/\tau(\omega)$ increases approximately linearly as a function of frequency, and when the temperature is increased, the $1/\tau(\omega)$ curves are shifted vertically proportional to T. The notion that $1/\tau(\omega,T)\sim\omega+T$ in the cuprates forms one of the center pieces of the Marginal Fermi Liquid model[1,13] and it has been shown to be approximately correct in a large number of experimental papers[10-12]. This phenomenology stresses the importance of temperature as the (only) relevant energy scale near optimal doping, which has motivated the idea that optimally doped cuprates are close to a quantum critical point[1]. As can be seen in Fig. 1, $1/\tau(\omega)$ has a negative curvature in the entire infrared region for all temperatures, and it saturates at around 5000 cm$^{-1}$. Although this departure from linearity may seem to be a minor detail, we will see that it is a direct consequence of the quantum critical scaling of the optical conductivity.

If a QPT indeed occurs at optimal doping $x=x_c$, then three major frequency regimes of qualitatively different behavior are expected[2]: (1) $\omega<T$; (2) $T<\omega<\Omega$; (3) $\Omega<\omega$. As we now report, we find direct indications of these regimes in our optical conductivity data.

**Region 1** ($\omega < T$) corresponds to measurement times long compared to the compactification radius of the imaginary time, $L_T = \hbar/k_BT$ (see **Methods**). Some ramifications have already been discussed above. In addition Sachdev[2] showed that in this regime the system exhibits a classical relaxational dynamics characterized by a relaxation time $\tau_r=AL_T$ (A is a numerical prefactor of order 1), reflecting that temperature is the only scale in the system. Assuming a Drude form, $\sigma_1(\omega)=(4\pi)^{-1}\omega_{pr}^2\tau_r/(1+\omega^2\tau_r^2)$, for the low frequency regime, $T\sigma_1(\omega,T)$ becomes a universal function of $\omega/T$, at least upto a number of order one:

$$\frac{\hbar}{k_B T \sigma_1(\omega,T)} = \frac{4\pi}{A\omega_{pr}^2}\left(1 + A^2\left(\frac{\hbar\omega}{k_B T}\right)^2\right) \qquad (1)$$

In the inset of Fig.2 we display $\hbar/(k_BT\sigma_1)$ as a function of $u=\hbar\omega/k_BT$. Clearly the data follow the expected universal behaviour for u < 1.5, with A=0.77. The experimental data are in this regard astonishingly consistent with Sachdev's predictions, including A≈1. From the fitted prefactor we obtain that $\omega_{pr}/2\pi c$ = 9597cm$^{-1}$. Above we have already determined the total spectral weight of the free carrier response, $(\omega_p/2\pi c)^2 = 19364^2$ cm$^{-2}$. Hence the classical relaxational response contributes 25% of the free carrier spectral weight. These numbers agree with the results and analysis of Quijada *et al*[10]. This spectral weight collapses into the condensate peak at $\omega=0$ when the material becomes superconducting[10]. In Fig. 2 we also display the scaling function proposed by Prelovsek[14], $\sigma_1(\omega)=C(1-\exp(-\hbar\omega/k_BT))/\omega$. The linear frequency dependence of this formula for $\hbar\omega/k_BT<<1$ is clearly absent from the experimental data. The universal dependence of $T\sigma_1(\omega,T)$ on $\omega/T$ also contradicts the "cold spot model"[15], where $T\sigma_1(\omega,T)$ has a universal dependence on $\omega/T^2$.

In **region 2** ($T < \omega < \Omega$) we can probe directly the scale invariance of the quantum critical state. Let us now introduce the scaling relation along the time axis, as follows from elementary consideration. It is standard knowledge that the Euclidean (i.e., imaginary time) correlator has to be known in minute detail in order to enable the analytical continuation to real (experimental) time. However, in the critical state invariance under scale transformations fixes the functional form of the correlation function completely: It has to be an algebraic function of imaginary time. Hence, it is also an algebraic function of Matsubara frequency $\omega_n=2\pi n/L_T$, and the analytical continuation is unproblematic: (i) Scale invariance implies that $\sigma_1(\omega)$ and $\sigma_2(\omega)$ have to be algebraic functions of $\omega$, (ii) causality forces the exponent to be the same for $\sigma_1(\omega)$ and $\sigma_2(\omega)$, and (iii) time reversal symmetry, implying $\sigma(\omega)=\sigma^*(-\omega)$, fixes the absolute phase of $\sigma(\omega)$. Taken together

$$\sigma(\omega) = C(-i\omega)^{\gamma-2} = C\omega^{\gamma-2}e^{i\pi(1-\gamma/2)} \qquad (2)$$

*Hence the phase angle relating the reactive and absorptive parts of the conductivity, $\arctan(\sigma_2/\sigma_1)=(2-\gamma)\cdot 90°$, is frequency independent and should be set by the critical exponent $\gamma$.* Powerlaw-behaviour of the optical conductivity of the cuprates has been reported previously[16,17], but to our knowledge the relation between the phase and the exponent has not been addressed in the literature. In Fig. 3 we display the frequency dependence of $|\sigma|$ in a log-log plot, and the phase in a linear plot. Although the temperature dependence leaks out to surprisingly high frequency in the latter, the data are remarkably consistent with Eq. 2 for $\omega$ between $k_BT$ and 7500 cm$^{-1}$. The observed powerlaw of the conductivity, $|\sigma|=C/\omega^{0.65}$ corresponds to $\gamma=1.35$, and the value of the phase, $\arctan(\sigma_2/\sigma_1)=60°\pm 2°$, implies that $\gamma=1.33\pm 0.04$. The good consistency of $\gamma$ obtained from two experimental quantities (i.e., the exponent of a powerlaw and the phase) is a strong test of the validity of Eq. 2. Frequency independence of the phase in region 2 and agreement between the two powerlaws (one from $\sigma(\omega)$ and the other from the phase angle spectrum) are unique properties of slightly overdoped samples, as demonstrated by Fig. 4, where we present the phase function for optimally doped[18] ($T_c$ = 88 K), underdoped[18] ($T_c$=66 K), and overdoped ($T_c$ = 77 K) single crystals of $Bi_{2.23}Sr_{1.9}Ca_{0.96}Cu_2O_{8+\delta}$ with different oxygen concentrations[19]. The observed trend for different dopings suggests that optimal doping, with $T_c$ = 88 K, and overdoping, with $T_c$ = 77 K, are lower and upper carrier concentrations where the optical conductivity obeys Eq. 2.

Because the phase is constant, the frequency dependent scattering rate $1/\tau(\omega)=\mathrm{Re}\{\omega_p^2/4\pi\sigma(\omega)\}=C\omega^{2-\gamma}=C\omega^{0.65}$ can *not* be a linear function of frequency (but note that $1/\tau^*(\omega)$ *is* linear[20], see **Methods**). Our findings disqualify directly theories that do not incorporate a manifest temporal scale invariance. Luttinger liquids are quantum critical states of matter and Anderson's results[20] based on one-dimensional physics are therefore of the correct form, Eq. 2. The exponent $\gamma=4/3$ is within the range considered by Anderson, but differs significantly from the prediction based on the "cold spot" model[15], providing $\sigma(\omega)\sim(-i\omega)^{-0.5}$, which corresponds to $\gamma=3/2$.

Let us now turn to the temperature dependence of the optical conductivity. From Fig. 2 we see, that in this region the conductivity crosses over to a different dependence on $\omega$ and T: For $\hbar\omega/k_BT>3$ the experimental data are seen to collapse onto a curve of the form $\sigma_1(\omega,T)=T^{-0.5}h(\omega/T)$ where $h(u)$ is a universal function. For $u>3$, $h(u)$ has a weak powerlaw dependence corresponding to a frequency dependence $\sigma_1(\omega,T)\sim\omega^{-0.65}$. According to the simplest scaling hypothesis, $\sigma_1(\omega,T)\sim T^{-\mu}h'(\omega/T)$ with $h'(u)\to$ constant and $h'(u)\to u^{-\mu}$ in the limits $u\to 0$ and $u\gg 1$, respectively. Although this energy-temperature scaling is roughly satisfied in the high frequency regime, it is strongly violated at low frequencies, because the success of Eq. 1 in the regime for $\hbar\omega/k_BT<1.5$ (see Fig. 2) implies an exponent $\mu=1$ at low frequencies instead. Bernhoeft has noticed a similar problem in the context of the heavy fermion critical points[21].

**Region 3** ($\omega>\Omega$) necessarily has a different behaviour of the optical conductivity, based on the following simple argument: The spectral weight of the optical conductivity integrated over all frequencies is set by the f-sum rule. However, since $\gamma>1$, the integration over all frequencies of Re$\sigma(\omega)$ of the form of Eq. 2 diverges. Hence we expect a crossover from the constant phase angle Arg$\sigma(\omega)= (2-\gamma)\cdot 90°$ to the asymptotic value Arg$\sigma(\infty)=90°$. The details of the frequency dependence of $\sigma(\omega)$ at the crossover point $\Omega$ depend on the microscopic details of the system. A (non-universal) example of an ultraviolet regularization with the required properties is[22] $\sigma(\omega)=(ne^2/m)(-i\omega)^{\gamma-2}(\Omega-i\omega)^{1-\gamma}$. Indeed the phase functions show a gradual upward departure from the plateau value for frequencies exceeding 5000 cm$^{-1}$ (Fig. 3). This indicates that the 'ultra-violet' cutoff is on the order of 1 eV.

Do our observations shed light on the enigmatic origin of the quantum criticality? In fact, they point unambiguously at three surprising features. *Firstly*, the current correlator behaves singularly and this implies that the electromagnetic currents themselves are the order parameter fields responsible for the criticality. *Secondly,* the criticality exceeds to surprisingly high energies. *Thirdly*, we have seen, that the optical conductivity curves collapse on $\sigma_1(\omega,T)=T^{\mu}h(\omega/T)$, where $\mu=1$ for $\omega/T<1.5$, while $\mu\sim 0.5$ for $\omega/T>3$. This disqualifies many theoretical proposals. Much of the intuition regarding quantum criticality is based on the rather well understood quantum phase transitions in systems composed of bosons. A canonical example is the insulator-superconductor transition in 2 space dimensions[23] where the optical conductivity is found to precisely obey the energy-temperature scaling hypothesis[24], characterized by a single exponent $\mu=0$ governing both the frequency and temperature dependences[2,24,25]. Bosonic theory can be therefore of relevance in electron systems but it requires that the fermionic degrees of freedom are bound in collective bosonic degrees of freedom at low energy. In the cuprates it appears that the quantum criticality has to do with the restoration of the Fermi-liquid state in the overdoped regime characterized by a large Fermi-surface. This implies that fermionic fluctuations play a central role in the quantum-critical state and their role has not yet been clarified theoretically. The absence of a single mastercurve for all values of ω/T is at variance with notions of quantum critical behaviour and its understanding may require concepts beyond the standard model of quantum criticality. We close with the speculation that the presence of bosonic fluctuations *and* fermionic fluctuations in the cuprates is pivotal in understanding the quantum critical behaviour near optimal doping of the cuprates.

**Methods**

**Kubo formula**

The Kubo formalism establishes the relation between optical conductivity and current-current correlation function

$$\sigma(i\omega_n, T) = \frac{Ne^2}{m\omega_n} + \frac{1}{\omega_n}\int_0^{L_T} d\tau\, e^{i\omega_n\tau} \langle \mathbf{j}(\tau)\mathbf{j}(0)\rangle$$

The first term, corresponding to perfect conductivity, is only relevant in the superconducting state, $\mathbf{j}(\tau)$ is the current operator at (imaginary) time $\tau$, while in the path integral formalism $L_T=\hbar/k_B T$ is the compactification radius of the imaginary time. The angle brackets mean a trace over the thermal distribution at finite temperatures. The integration is over a finite segment of imaginary time, resulting in the optical conductivity at the Matsubara frequencies $i\omega_n=2\pi in/L_T$ along the imaginary axis of the complex frequency plane. The conductivity at real frequency ω follows from analytical continuation.

**Experimental determination of the optical conductivity**

The most direct experimental technique, which provides the optical conductivity and its phase, is spectroscopic ellipsometry. Another popular approach is the measurement of the reflectivity amplitude over a wide frequency region. Kramers-Kronig relations then provide the phase of the reflectivity at each frequency, from which with the help of Fresnel equations the real and imaginary part of the dielectric function, ε(ω), is calculated. We used reflectivity for 50 cm$^{-1}$ < ω/2πc < 6000 cm$^{-1}$, and ellipsometry for 1500 cm$^{-1}$<ω/2πc<36000 cm$^{-1}$. This combination allows a very accurate determination of ε(ω) in the entire frequency range of the reflectivity and ellipsometry spectra. Due to the off-normal angle of incidence used with ellipsometry, the ab-plane pseudo-dielectric function had to be corrected for the c-axis admixture. We used previously published[8] c-axis optical constants of the same compound. The data files have been generously supplied to us by S. Tajima. The effect of this correction on the pseudo-dielectric function turns out to be almost negligible, in accordance with Aspnes[9].

The optical conductivity, σ(ω), is obtained using the relation $\varepsilon(\omega)=\varepsilon_\infty+4\pi i\sigma(\omega)/\omega$, where $\varepsilon_\infty$ represents the screening by interband-transitions. In the cuprate materials $\varepsilon_\infty$=4.5±0.5. For ω/2πc=5000 cm$^{-1}$ an uncertainty of 0.5 of $\varepsilon_\infty$ propagates to an error of 2° degrees of the phase of σ(ω). This accuracy improves for lower frequencies.

**Frequency dependent scattering rate**

For an isotropic Fermi-liquid the energy dependent scattering rate of the quasi-particles can be readily obtained from the optical data, using the relation $1/\tau(\omega)=\text{Re}\{\omega_p^2/4\pi\sigma(\omega)\}$. In spite of the fact that the notion of a quasi-particle in the spirit of Landau's Fermi-liquid is far from being established for the cuprates, during the past one and a half decade it has become a rather common practice to represent infrared data of these materials as 1/τ(ω). The dynamical mass is defined as $m^*(\omega)/m=\text{Im}\{\omega_p^2/4\pi\omega\sigma(\omega)\}$. To obtain absolute numbers for 1/τ(ω) and $m^*(\omega)/m$ from the experimental optical conductivity, a value of the plasma-frequency, $\omega_p$, must be adopted. With our value of $\omega_p$ the dynamical mass converges to 1 for ω→∞. Sometimes the renormalized scattering rate, $\tau^*(\omega)^{-1}=\tau(\omega)^{-1}m/m^*(\omega)=\omega\sigma_1(\omega)/\sigma_2(\omega)$, is reported instead of τ(ω). If the frequency dependence of the conductivity is a power law, $\sigma(\omega)=(-i\omega)^{\gamma-2}$, then $1/\tau^*(\omega)=-\omega\cotan(\pi\gamma/2)$, which is a linear function of frequency[20]. The value of the slope reveals the exponent, and corresponds to the phase of the conductivity displayed in Figs. 3-4.

**Ackowledgements**. This investigation was supported by the Netherlands Foundation for Fundamental Research on Matter (FOM) with financial aid from the Nederlandse Organisatie voor Wetenschappelijk Onderzoek (NWO). The crystal growth work at Stanford University was supported by the Department of Energy's Office of Basic Energy Sciences, Division of Materials Science.We thank C. M. Varma, P. Prelovsek, C. Pepin, S. Sachdev, and A. Tsvelik for helpful comments during the preparation of this work, and N. Kaneko for technical assistence.

**Competing interests statement.** The authors declare that they have no competing financial interests.

**Correspondence and requests for materials should be addressed to** dirk.vandermarel@physics.unige.ch

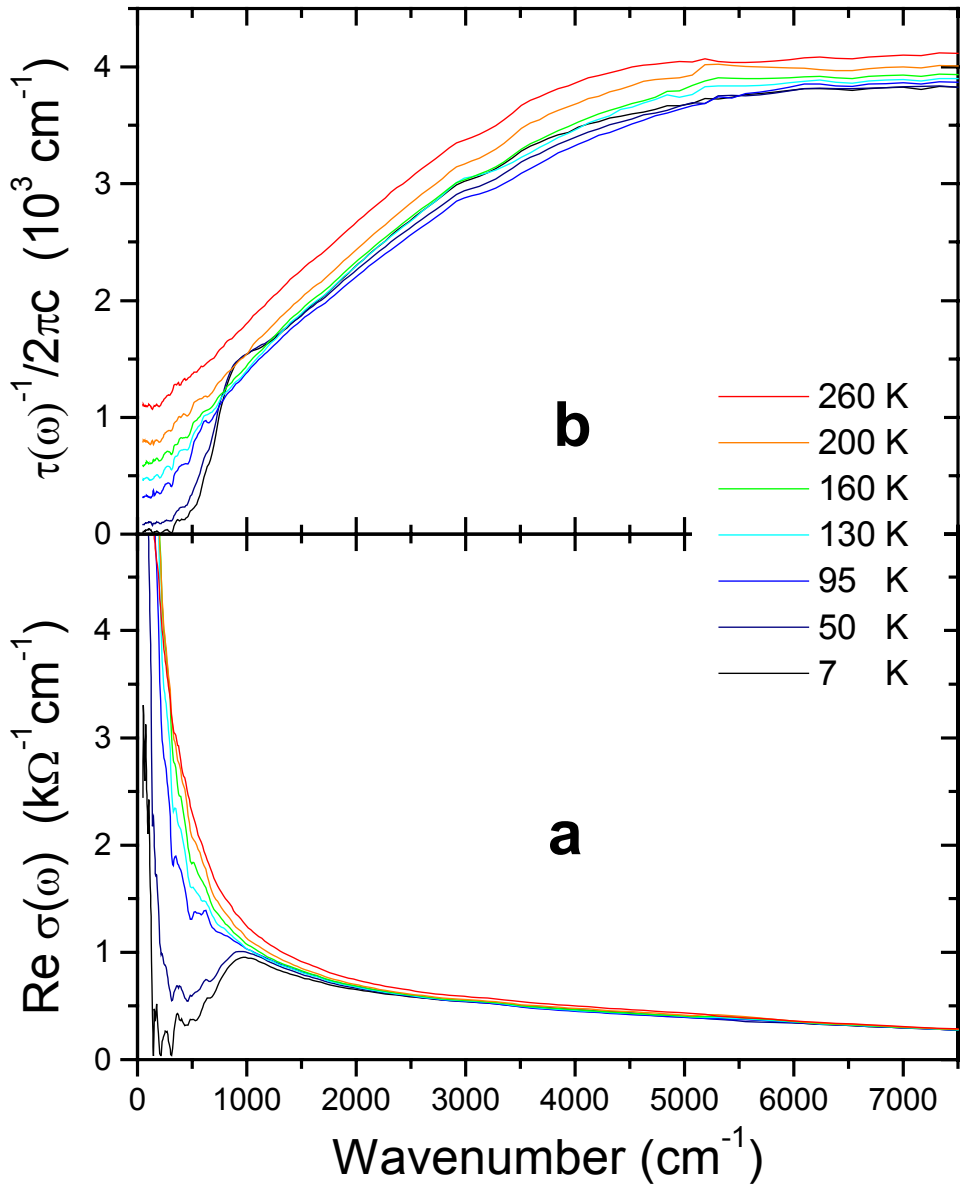

**Figure 1** Optical properties along the copper-oxygen planes of $Bi_2Sr_2Ca_{0.92}Y_{0.08}Cu_2O_{8+\delta}$ for a selected number of temperatures. (a) Optical conductivity and (b) the frequency dependent scattering rate defined as $1/\tau(\omega)=Re\{\omega_p^2/4\pi\sigma(\omega)\}$ (see Methods). The relatively high transition temperature ($T_c$=96 K) of this crystal compared to previous reports on Bi-2212 is caused by the partial substitution of yttrium on the calcium sites.

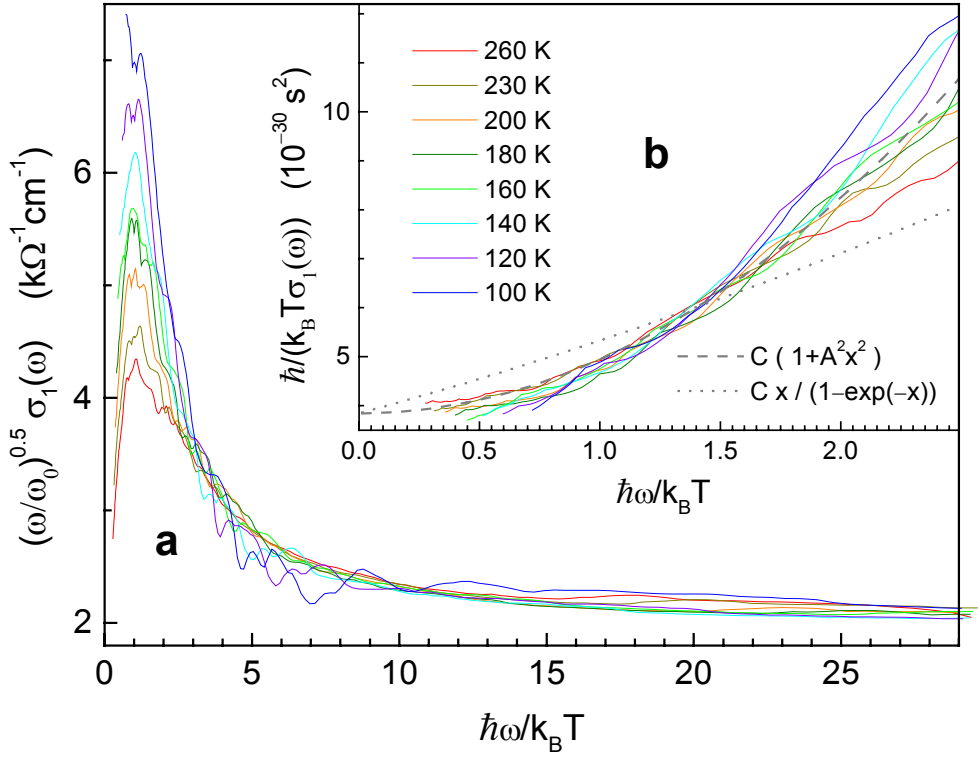

**Figure 2** Temperature/frequency scaling behaviour of the real part of the optical conductivity of $Bi_2Sr_2Ca_{0.92}Y_{0.08}Cu_2O_{8+\delta}$. The sample is the same as in Fig. 1. In a the data are plotted as $(\omega/\omega_0)^{0.5}\sigma_1(\omega,T)$. The collapse of all curves on a single curve for $\hbar\omega/k_BT>3$ demonstrates that in this $\omega/T$-region the conductivity obeys $\sigma_1(\omega,T)=\omega^{-0.5}g(\omega/T)=T^{-0.5}h(\omega/T)$. Note that $g(u)=u^{0.5}h(u)$. In panel b the data are presented as $\hbar/(k_BT\sigma_1(\omega,T))$, demonstrating that for $\hbar\omega/k_BT<1.5$ the conductivity obeys $\sigma_1(\omega,T)=T^{-1}f(\omega/T)$.

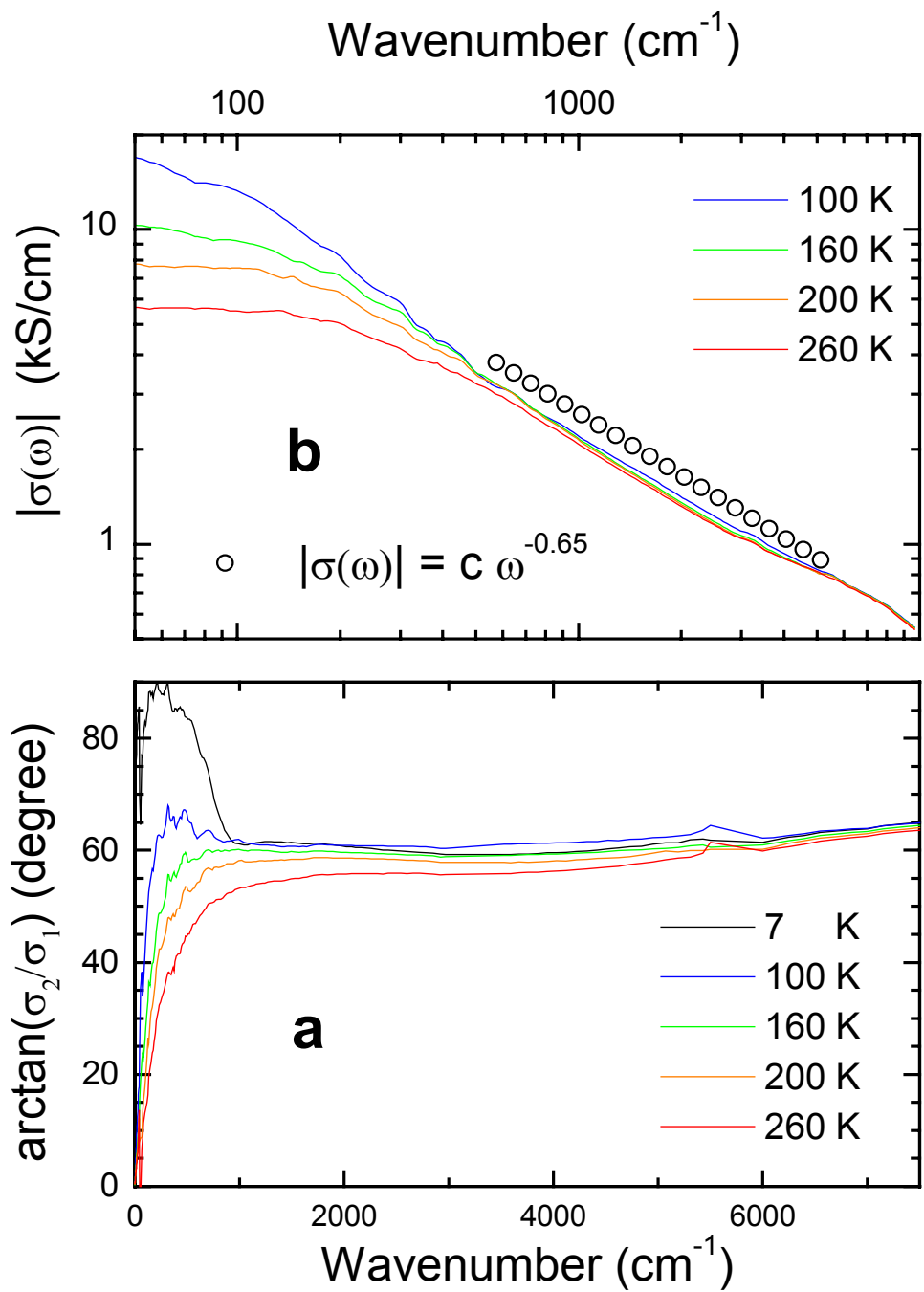

**Figure 3** Universal powerlaw of the optical conductivity and the phase angle spectra of optimally doped $Bi_2Sr_2Ca_{0.92}Y_{0.08}Cu_2O_{8+\delta}$. The sample is the same as in Fig. 1. In a the phase function of the optical conductivity, $Arg(\sigma(\omega))$ is presented. In b the absolute value of the optical conductivity is plotted on a double logarithmic scale. The open symbols correspond to the powerlaw $|\sigma(\omega)|=C\omega^{-0.65}$.

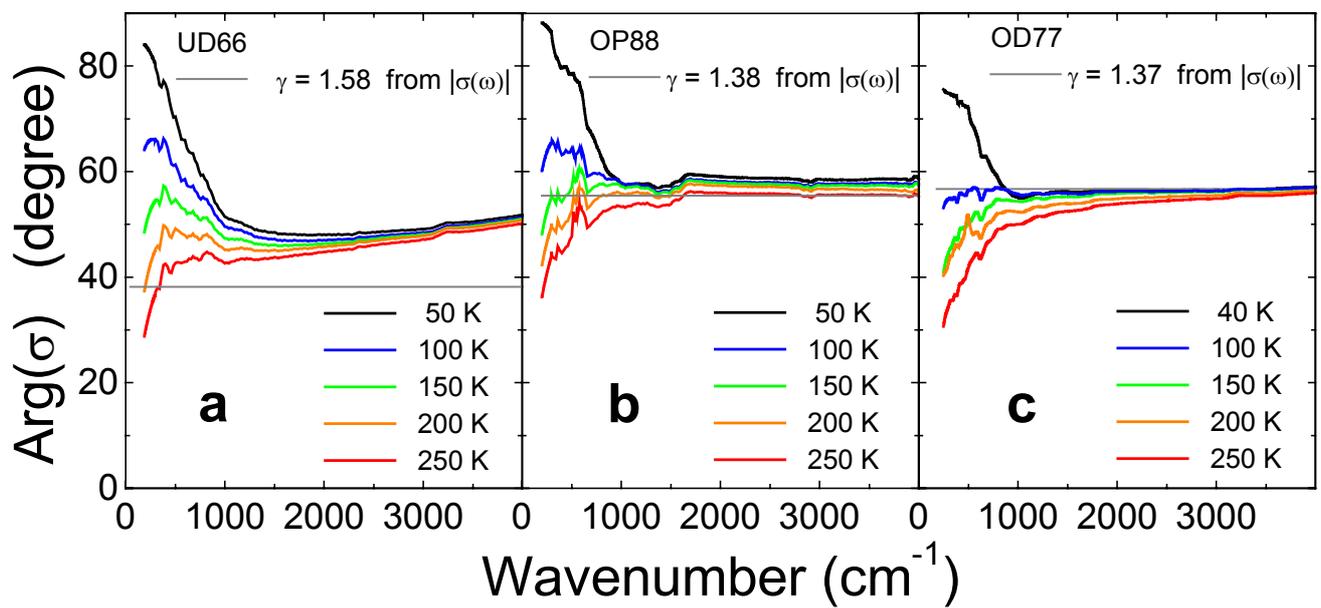

**Figure 4** Phase of $\sigma(\omega)$ of underdoped ($T_c$=66 K, panel a), optimally doped ($T_c$=88 K, panel b) and overdoped ($T_c$=77 K, panel c) $Bi_{2.23}Sr_{1.9}Ca_{0.96}Cu_2O_{8+\delta}$. Solid lines: exponent from $|\sigma(\omega)|$ between 1000 and 5000 cm$^{-1}$. These crystals were grown at relatively low partial pressures (25 mbar) of oxygen, resulting in high-quality underdoped crystals with $T_c$ as low as 65K, and with sharp superconducting transitions $\delta T_c \leq 2.5$ K. The optimally doped and overdoped crystals were obtained by post-annealing in oxygen. At optimal doping $T_c$ turns out to be somewhat lower than the highest values reported in Bi-2212 due to a slightly different cation ratio.